\documentclass[11pt,a4paper]{article}
\begin{document}
\thispagestyle{empty}
\newcommand{\PSI}[1]{\Psi_{\textrm{\footnotesize{#1}}}}
\newcommand{\PHI}[1]{\Phi_{\textrm{\footnotesize{#1}}}}
\newcommand{\sta}[2]{\psi^{(#2)}\!(#1)}
\newcommand{\state}[2]{\psi^{(#2)}\!(#1)}
\newcommand{\statex}[2]{\psi^{(#2)}\!(#1,m,\chi_{#1})}

\newcommand{\statexx}[3]{\psi^{(#3)}\!(#1,m,\chi_{#2})}
\newcommand{\staexx}[3]{\psi^{(#3)}\!(#1,{#2})}
\newcommand{\staexxm}[3]{\psi^{(#3)}\!(#1,m_{#2})}
\newcommand{\cO}     {c_{\scriptscriptstyle {\mathrm{O}}}}
\newcommand{\sbl}{\hspace{1pt}}
\newcommand{\stat}[1]{\psi(#1,m)}
\newcommand{\bfitr}{\emph{\boldmath $r$}}
\newcommand{\et}{\mbox{$\eta_{21}$}}
\newcommand{\per}{\mbox{$P_{21}$}}
\newcommand{\palf}{$P_{\alpha}\;$}
\newcommand{\rmi}{\mathrm{i}}
\noindent
{\Large \textbf{Connecting spin and statistics in quantum mechanics}}\footnote{An earlier version has been paper-published in Found. Phys. {\bf40} (7), 776-792, 793-794 (2010).}
\\
\\
{\bf{Arthur Jabs}}
\\
\\
Alumnus, Technical University Berlin

\noindent
Vo\ss str. 9, 10117 Berlin, Germany

\noindent
arthur.jabs@alumni.tu-berlin.de
\\
\\
(3 February 2014)
\bigskip

\noindent
{\bf{Abstract.}} The spin-statistics connection is derived in a simple manner under the postulates that the original and the exchange wave functions are simply added, and that the azimuthal phase angle, which defines the orientation of the spin part of each single-particle spin-component eigenfunction in the plane normal to the spin-quantization axis, is exchanged along with the other parameters. 
The spin  factor $(-1)^{2s}$ belongs to the exchange wave function when this function is constructed so as to get the spinor ambiguity under control. This is achieved by effecting the exchange of the azimuthal angle by means of rotations and admitting only rotations in one sense. The procedure works in Galilean as well as in Lorentz-invariant quantum mechanics. Relativistic quantum field theory is not required.
\vspace{6pt}
\begin{list}
{\textbf{Keywords:}}
{\setlength{\labelwidth}{2.0cm} \setlength{\leftmargin}{2.2cm}
\setlength{\labelsep}{0.2cm}}
\item spin and statistics, spinor, spinor ambiguity, Bose and Fermi statistics, Pauli exclusion principle, symmetrization 
\end{list}
\vspace{5pt}
\rule{\textwidth}{.3pt}
\newcommand{\leer}{\hspace*{\fill}}
\newcommand{\lll}{\hspace*{10pt}}
\newcommand{\hsp}{\hspace*{60pt}}
\begin{center}
\vspace{-20pt}
\noindent
 \hsp 1~~ Introduction \leer 1\hsp \lll \\
\hsp 2~~ The special parameter \leer 4\hsp \lll \\ 
 \hsp 3~~ Controlling the spinor ambiguity \leer 5\hsp \lll \\
\hsp 4~~ Constructing the exchange function \leer 6\hsp \lll \\
\hsp 5~~ Generalization to $N$ particles \leer 8\hsp \lll \\
 \hsp 6~~ Generalization to non-product wave functions \leer 9\hsp \lll \\
\hsp 7~~The relativistic domain \leer 9\hsp \lll \\
 \hsp Notes and references	\leer 10-13\hsp \lll \\
\end{center}
\vspace{-10pt}
\noindent
\rule{\textwidth}{.3pt}
\vspace{-8pt}
\begin{flushright}
\emph{Physics is simple but subtle}

Paul Ehrenfest
\end{flushright}
{\textbf{1~~Introduction}}

\noindent
The standard method of treating systems of identical particles in present quantum mechanics is to require that every wave function or state vector must be either symmetric
or antisymmetric, that is, multiplied by either $+1$ or $-1$ when the labels or parameters referring to any two particles are interchanged. There are thus two classes of systems, with different collective behaviour of the particles: systems of bosons and systems of
fermions. These two classes are connected with the spins of the particles: all particles 
which are known to be bosons are empirically found to have integral spin, in units of $\hbar$, while all known fermions 
have half-integral (i.e., half-odd-integral) spin.

	Within quantum mechanics the connection with spin could not be derived and had to be taken as another postulate. The 
first derivation was provided by Fierz [1] and Pauli [2], who founded it on relativistic quantum field theory. 
This also remained the framework for the papers which in subsequent years refined and
generalized Pauli's proof [3, 4]. Typically, in these papers it is postulated that no negative-energy states exist, that the metric in Hilbert space is positive definite, and that the fields either commute or anticommute for spacelike separations (locality, microcausality). Under these conditions it is shown that integral-spin fields cannot satisfy the (fermionic) anticommutation relations, and half-integral-spin fields cannot satisfy the (bosonic) commutation relations. This does not exclude the possibility that fields exist which satisfy other commutation relations and show statistics that differ from Bose and Fermi statistics.

In 1965 Feynman in his Lectures on Physics [5] objected:
\begin{quote}
	An explanation has been worked out by Pauli from complicated arguments of quantum field theory and relativity. He has shown that the two [spin and statistics] must necessarily go together, but we have not been able to find a way of reproducing his arguments on an elementary level. It appears to be one of the few places in physics where there is a rule 
which can be stated very simply, but for which no one has found a simple and easy explanation. The explanation is deep down in relativistic quantum mechanics. This probably means that we do not have a complete understanding of the fundamental principle involved.
\end{quote}
The aim of the present paper is to propose such a simple and easy explanation.

Actually, since 1965 more than a hundred publications appeared deriving the spin-statistics connection under different sets of conditions [6]. Reviews are contained in [7--10]. Many of these publications derive the connection in settings far removed from standard (local) relativistic quantum field theory; and they are also far from simple and easy.

Closest to the present approach are those papers that use only quantum mechanics, relativistic or nonrelativistic, and are written in the spirit of Feynman's demand for simplicity. These papers nevertheless contain one or several of the following restrictions: the wave functions must have special invariance [11, 12], continuity [13--17] or symmetry [18] properties, or   must lie in special spin-component subspaces [19]. The systems considered must be nonrelativistic [13--22], have only two spatial dimensions [22], contain only two particles [18, 19], only particles with zero spin [13--17] or spin $\leq 1/2$ [20, 21], only point particles [23--25], must admit antiparticles [26], or the exchange must be considered as physical transportation of real objects [11, 12, 23--25, 27].

The present proposal is not subject to any of these restrictions. It grew out of an attempt to understand the papers by York [28, 29] in the framework of the realist interpretation which I developed some time ago [30, 31]. The premises of the present proposal are seen when the organization of the paper is considered. 

We start with a proposal by Feynman made in an attempt to derive the spin-statistics connection. In [27] Feynman suggests that we may
\begin{quote}
take the view that the Bose rule is obvious from some kind of understanding that the amplitude[s] in quantum mechanics that correspond to alternatives must be added.
\end{quote}
We follow his proposal in that, in the construction of a symmetric or antisymmetric wave function for a system of identical particles, we just add up the original and the exchange function. The exchange function is defined as the original wave function in which the labels or parameters referring to the single particles have been exchanged. That is, in the case of two particles we start from
\begin{equation}
\Psi_{\textrm {S}}=\frac{1}{\sqrt{2}}\Big(\psi(1,2)+\psi(2,1)\Big)
\end{equation}
and in the case of $N$ particles from
\begin{equation}
\Psi_{\textrm {S}}=\frac{1}{\sqrt{N!}}\left(\sum\nolimits_{\alpha}P_{\alpha}\psi(1,2,\ldots,N)\right),
\end{equation}
where $P_{\alpha}$ is a permutation of the parameters referring to the single particles. The sum (2) extends over the $N!$ possible permutations, including the identity $I$. It is the extension of the sum (1) from two-particle to $N$-particle systems and corresponds to the totally symmetric function in standard quantum mechanics.

We shall see that the minus sign in the superposition of fermionic wave functions arises from the construction of the exchange function. In Sect.~2 we point out that it is important to consider the azimuthal spin angle $\chi$, which defines the orientation of the spin part of a single-particle spin-component eigenfunction. The angle $\chi$ is also exchanged, but requires a special treatment because it expresses the well known spinor ambiguity when the spin component $m$ is half-integral. This means that we cannot know which of the two possible values of the function with the exchanged $\chi$ has to be chosen.

In Sect.~3 it is shown that the ambiguity can be overcome by effecting the exchange of $\chi$ by way of rotations and by admitting only rotations in one direction, either clockwise or counterclockwise. In Sect.~4 the rotations leading from the original to the exchange function are explicitly carried out, and it is shown that the exchange function thereby acquires the desired spin factor $(-1)^{2s}$. It is thus the rotation group, a subgroup of both the Galilei and the Lorentz group, that determines the type of statistics. In standard quantum mechanics it is the permutation group that does this: its one-dimensional representations are associated with Bose and Fermi statistics, and its other representations with ``parastatistics''. The rotation group in our approach leads only to Bose and Fermi statistics, and there is no reason to suggest experiments in search of particles with parastatistics.

In Sect.~5 the proof is extended to $N$ particles, and in Sect.~6 to non-product wave functions. Finally, in Sect.~7, using the properties of helicity functions, it is pointed out that the proof also holds in relativistic quantum mechanics.
\newline

\noindent
{\textbf{2~~The special parameter}

\noindent
In order to present the essential points in a simple way we begin by considering a non-relativistic (Galilean) system of two identical particles of spin $s$ ({\emph{\boldmath$S$}}$^2\,\psi = s(s+1)\,\psi$) described by a (Schr\"odinger) wave function which is a product of two normalized one-particle wave  functions

\begin{equation}
\Psi = \state{a,m_a}{1} \;\state{b,m_b}{2}\, ,
\end{equation}
and the one-particle wave functions are eigenfunctions of the operator of the spin component with respect to an arbitrary but common spin-quantization axis. These restrictions will be removed in Sects.~5 to 7.
 
The single-particle wave functions are functions of the variables $x, y, z, t$. The functions are determined by the parameters $a, b, m_a, m_b$, where $a$ and $b$ stand for the sets of parameters that, together with $m_a,m_b$ and $\chi_a,\chi_b$ (below), allow for a \emph{complete} account of all aspects and degrees of freedom of the single-particle systems. Such a set of parameters includes e.g. mass, charge, total spin, centre, expansion coefficients etc. Mass, charge, and total spin are of course the same for identical particles and their exchange has no effect. An alternative notation would be $\psi(a,m_a,x^{(1)},y^{(1)},z^{(1)},t)$, where the labels in parentheses, (1) and (2), distinguish the particles in the formalism. We have suppressed here the variables and have put the particle labels directly at the function symbols. 

Among the parameters of the wave function there is one that requires special treatment in the construction of the exchange function because it may lead to double-valued wave functions. The reason is that the spin parts of the wave functions, while belonging to one and the same $m$, may still differ from one another by a rotation about the spin-quantization axis. In other words, each spin part has a definite orientation in a plane normal to the common spin-quantization axis, defined by an azimuthal angle $\chi$, counted from some arbitrary reference direction. A complete spin-quantization \emph{frame} rather than only a spin-quantization \emph{axis} is involved. The angle $\chi$ is kept out of the set $a,m_a$ (and $b,m_b$) and is exhibited explicitly. 
Each function can have its own angle, but the particular values do not matter. The values of $\chi$ are restricted to the interval $[0,2\pi]$.

The specific form of the parametric dependence of the spin-component eigenfunction on $\chi$ is given by the factor
\begin{equation}
\exp(\rmi m \chi),
\end{equation}
so that we extend (3) to read
\begin{equation}
\Psi=\exp(\rmi m_a \chi_a)\,\sta{a,m_a}{1}\;\exp(\rmi m_b \chi_b)\,\sta{b,m_b}{2}. 
\end{equation}
The parameter $m$ appears twice; in the exponential it governs the dependence on $\chi$, and in the list of arguments of $\psi$ it serves to mark the component of the spin vector even when $\chi$ is zero or is omitted.

The angle $\chi$ appears ``only'' in a phase factor, and when this factor is an overall (global) phase factor it is without any physical significance and can be omitted. However, it must be taken seriously if it is to become part of a superposition, thereby determining a \emph{relative} phase and thus becoming physically significant [32, pp. 219, 220]. This is what happens in the present approach: the angle $\chi$ is exchanged, though in a specific way, along with the other parameters, and in Sect.~4, when the original and the exchange function are superposed, $\chi$ becomes instrumental in determining the relative phase between these functions.

The exponential factor (4) expresses the spinor ambiguity: in the case of half-integral $m$ it is $+1$ for $\chi=0$ and $-1$ for $\chi=2\pi$.
The sign change under a full rotation holds for every rotation axis. The factor (4) refers to the particular case of rotations about the spin-quantization axis. This does not lead out of the subspace of functions with the same $m$. All this is standard quantum mechanics [32, pp. 694, 703, 985, 986].

Note that the rotation of the \emph{orbital} angular momentum part of a wave function can be expressed by changing the value of an already present spatial variable ($\varphi\rightarrow\varphi+\varphi'$ [32, pp. 681, 699]), but that an additional parameter ($\chi$) is needed to express the behaviour of the \emph{spin} part of a wave function under a rotation.

The letter $\chi$, rather than the customary $\varphi$, is used in order to emphasize that it is the spin part, not the orbital part, which is concerned, that the spin-quantization axis need not coincide with the $z$-axis, and that the angle $\chi$ is not a variable of the one-particle wave function, as $\bfitr$ and $\varphi$ are. Rather, the dependence on $\chi$ is a parametric dependence, like that on $m$ and the other parameters in $a$ and $b$. Therefore the application of a differential operator like $-\rmi \hbar\partial/\partial\chi$, analogous to the $z$-component of \emph{orbital} angular momentum, does not make sense for the spin part of the wave function.

In constructing the exchange wave function in traditional quantum mechanics it is irrelevant whether we exchange the particle labels (1), (2) or the function parameters $a, b,m_a,m_b,\chi_a,\chi_b$. In our construction it is no longer irrelevant, and it is the exchange of the parameters that must be chosen. Thus we replace $a,m_a$ by $b,m_b$ and vice versa in the original wave function (5). But because the parameter $\chi$ expresses the spinor ambiguity the exchange of $\chi_a$ with $\chi_b$ and vice versa cannot be done in such a simple way.
\newline

\noindent
{\textbf{3~~Controlling the spinor ambiguity}

\noindent
The special feature with the factor (4) is that $\chi$ is an angle, so that we may go from some particular value $\chi_a$ to some other value $\chi_b$ in two ways, either clockwise or counterclockwise. In the case of half-integral $m$ one way leads to a different wave function at $\chi_b$ than the other, the two functions having different signs. In other words, the value of the function at $\chi_b$ then depends not only on the value of $\chi_b$ but also on the path leading from $\chi_a$ to $\chi_b$. This leads to double-valued functions and represents another aspect of the spinor ambiguity.

One may imagine the function $\exp(\rmi m \chi)$ with half-integral $m$ to lie on the two-sheeted Riemannian surface of the function $\sqrt z$ [33, 34], where one sheet carries only one set of function values. The clockwise path from $\chi_a$ to $\chi_b$ always ends up in a different sheet than the counterclockwise path. Or one may imagine a M\"obius band, where on the first round trip over the band one set of function values is met, and the corresponding other set on the second round trip. In fact, devices like twisted ribbon belts [27, p.~58], contortions of an arm holding a cup [27, p.~30] and others [35--40] are similar to the Riemannian surface and the M\"obius band in that they construct an indicator of whether we are in the first or in the second turn, and in that they return to the original situation after the second turn. For integral $m$ (including $s=m=0$) the Riemannian surface has only one sheet and no ambiguity arises.

Now, when adding the original and exchange wave functions the functions must be uniquely defined. This is not the same as the general requirement that wave functions be single-valued. Single-valuedness can only be required for meas\-urable quantities such as transition probabilities or expectation values, but not for the wave functions themselves [41]. In many textbooks it is nevertheless invoked for the wave functions themselves, in particular for justifying the restriction to integral values of $m$ for orbital angular momentum. The real justification of integral $m$ here rests on group representations and properties of observables [42].

Our case is different because we are concerned with the procedure of constructing one wave function by superposition of others, formally similar to interference. 

Now, according to what has been said above the spinor ambiguity is removed (i.e., kept under control) if we make a choice between the two possible paths from $\chi_a$ to $\chi_b$, that is, if we exchange the $\chi$\sbl s by way of rotations and decide to make all rotations in one sense only, either clockwise or counterclockwise.

In the language of group theory the clockwise and the counterclockwise way from ${\chi}_a$ to $\chi_b$ correspond to paths of different homotopy classes (e.g. [43]). So our choice means that we are admitting only paths of the same homotopy class.
\newline

\noindent
{\textbf{4~~Constructing the exchange function}

\noindent
We are now ready to take the decisive step. We want to construct the exchange function from the original function (5), not by simply replacing $\chi_a$ by $\chi_b$ in the wave function $\,\exp(\rmi m_a\chi_a)\psi^{(1)}$  and $\chi_b$ by $\chi_a$ in the wave function $\,\exp(\rmi m_b\chi_b)\psi^{(2)}$ (as is done with the other parameters), but by continuously rotating the spin part of the functions from $\chi_a$ to $\chi_b$ and from $\chi_b$ to $\chi_a$ respectively, with due consideration being given to the paths connecting $\chi_a$ and $\chi_b$.

Thus, we start from formula~(5) where the $a,m_a$ and $b,m_b$ have already been exchanged, but the $\chi$\sbl s have not:
\begin{equation}
\Psi=\exp(\rmi m_b \chi_a)\;\sta{b,m_b}{1}\,\times\,\exp(\rmi m_a \chi_b)\;\sta{a,m_a}{2}.
\end{equation}
We then rotate the first function in (6) from $\chi_a$ to $\chi_b$. We take the counterclockwise sense of the rotations, and we assume $\chi_a < \chi_b$ and $m_a,m_b\geq 0$. In order to get from ${\chi}_a$ to ${\chi}_b$ we then have to run through ${\chi}_b - {\chi}_a$. This yields the rotation factor 
$\exp({\rmi m_b(\chi_b - \chi_a)}) $ and the first function turns into 
\begin{equation}
\exp(\rmi m_b\chi_b) \staexx{b}{m_b}{1}.
\end{equation}
Likewise, rotating the second function in (6) counterclockwise from $\chi_b$ to $\chi_a$ means that we have to run through \ ${2\pi-(\chi}_b - {\chi}_a)$. This yields the rotation factor \ $\exp({\rmi m_a(2\pi+\chi_a-\chi_b)})$ \  and the second function turns into
\begin{equation}
\exp(\rmi m_a(2\pi+\chi_a)) \staexx{a}{m_a}{2}.
\end{equation}
Writing Eq.~(5) with the new factors (7) and (8) then yields the exchange function
\begin{equation}
F \times \exp(\rmi m_b\chi_b)\staexx{b}{m_b}{1}\;\exp(\rmi m_a \chi_a)
\staexx{a}{m_a}{2}
\end{equation}
with
\begin{equation}
F=\exp(\rmi m_a 2\pi) =(-1)^{2m_a}=(-1)^{2s}, \hspace{1.2 em}
\end{equation}
where for the last equality we have used the fact that $s$ and $m$ are either both integral or both half-integral. Had we chosen the clockwise sense we would have obtained $F=\exp({-\rmi m_a 2\pi})$, which is also equal to $(-1)^{2s}$. The same result obtains when $m_a,m_b<0$ or when $\chi_a > \chi_b$, or when first rotating the $\chi\sbl$s and then exchanging the $m\sbl$s. The case $\chi_a = \chi_b$ is of statistical weight zero and can be neglected.

Finally, adding the original function (5) and the exchange function (9) (with (10)) we arrive at
\begin{displaymath}
\PSI{S}=\frac{1}{\sqrt{2}}\Big(\exp(\rmi m_a \chi_a)\;\sta{a,m_a}{1}\;\exp(\rmi m_b \chi_b)\;\sta{b,m_b}{2}
\end{displaymath}
\begin{equation}
\hspace{20pt}+ (-1)^{2s}\exp(\rmi m_b \chi_b)\;\sta{b,m_b}{1}\;\exp(\rmi m_a \chi_a)\;\sta{a,m_a}{2}\Big). 
\end{equation}
The factor $\exp(\rmi(m_a\chi_a+m_b\chi_b))$ can be drawn out of the sum and is thus an overall phase factor in front of $\PSI{S}$, and there it can be omitted.

Thus we are returning to the standard form of the wave functions, which do not explicitly exhibit the dependence on $\chi$:
\begin{equation}
\PSI{S} = \frac{1}{\sqrt{2}}\Big(\sta{a,m_a}{1}\; \sta{b,m_b}{2} + (-1)^{2s}\, \sta{b,m_b}{1}\;\sta{a,m_a}{2}\Big).  
\end{equation}
There is some formal analogy with interference between two parts of a split wave. One part is left unmodified [wave function (5)], the other is subject to a phase shift [exchange, wave function (9)], and then the two are recombined [wave function (11) or (12)].

With this we have reached our goal for the considered class of functions: the factor $(-1)^{2s}$ is no longer postulated but is derived in a simple way from basic principles. This factor yields $+1$ (bosons) for integral $s$ and $-1$ (fermions) for half-integral $s$, and this is the desired connection between spin and statistics.
\newpage
 
\noindent
{\textbf{5~~Generalization to $N$ particles}

\noindent
We begin now to remove the restrictions imposed on the wave function in the previous sections. In the present section we remove the restriction to two particles. The $N$-particle functions considered here are still of product form,
\begin{equation}
\Psi = e^{\rmi m_1\chi_1}\staexx{u_1}{m_1}{1}\,\cdots\, e^{\rmi m_N\chi_N}\staexx{u_N}{m_N}{N},
\end{equation}
where $u_1,m_1,\chi_1,\cdots\;$ replace $a,m_a,\chi_a,\cdots\;$ of Sects.~2 to 4. The function, when symmetrized with respect to the $u\sbl$s and $m\sbl$s, becomes 
\begin{equation}
\PSI{S$'$}=\frac{1}{\sqrt{N!}}\sum\nolimits_{\alpha}P_{\alpha}\,e^{\rmi m_1\chi_1}\staexx{u_1}{m_1}{1}\,\cdots\, e^{\rmi m_N\chi_N}\staexx{u_N}{m_N}{N}. 
\end{equation}
The index S$'$ (with prime) is to indicate that 
$P_{\alpha}$ in (14) permutes the parameter sets $u_i,m_i$ among the one-particle functions but does not permute the angles $\chi_i$. The permutation of the angles will be effected separately, by way of rotations. As any permutation can be written as a product of a number of transpositions, the term $P_{\alpha}\,\staexx{u_1}{m_1}{1}\cdots \staexx{u_N}{m_N}{N}$ differs from the term with $P_{\alpha}=I$ by a number $k_{\alpha}$ of transpositions. When the $\chi$ rotations are applied, as described in the preceding sections for the case of two particles, every single transposition yields the factor $F=(-1)^{2s}$ in front of the term with interchanged parameters, independent of the angles $\chi$. Hence $k_{\alpha}$ transpositions yield the factor $(-1)^{2sk_{\alpha}}$. The function (14) then changes into the superposition function (symmetric or antisymmetric)
\begin{displaymath}
\PSI{S}=\frac{1}{\sqrt{N!}}\sum\nolimits_{\alpha}(-1)^{2sk_{\alpha}}P_{\alpha}\;e^{\rmi m_1\chi_1}\staexx{u_1}{m_1}{1}\,
\cdots\, e^{\rmi m_N\chi_N}\staexx{u_N}{m_N}{N} 
\end{displaymath}
\begin{equation}
=\frac{1}{\sqrt{N!}}\sum\nolimits_{\alpha}(-1)^{2sk_{\alpha}}P_{\alpha}\;e^{\rmi (m_1\chi_1+\cdots+m_N\chi_N)}\;\staexx{u_1}{m_1}{1}\,
\cdots\, \staexx{u_N}{m_N}{N}.
\end{equation}
The index S (without prime) instead of S$'$ (with prime, as in (14)) is to indicate that the exchange of the angles $\chi$ by means of rotation is included. In (15) $P_{\alpha}$ therefore permutes the sets $u_i,m_i,\chi_i$ among each other. The single functions $\;(-1)^{2sk_{\alpha}}P_{\alpha}\,\exp(\rmi m_1\chi_1)\:\staexx{u_1}{m_1}{1}\cdots$ $\exp(\rmi m_N\chi_N)\staexx{u_N}{m_N}{N}\;$ for $P_{\alpha}\neq I\;$ are the extensions of the exchange function from two to $N$ particles. For half-integral $s$ the sum (15) is the well-known Slater determinant, which leads to the Pauli exclusion principle.              

The exponential factor under the sum in (15) remains unchanged under the permutations and can thus be drawn out of the sum and be omitted as an overall phase factor in front of $\PSI{S}$. We are thus left with
\begin{equation}
\PSI{S}=\frac{1}{\sqrt{N!}}\sum\nolimits_{\alpha}(-1)^{2sk_{\alpha}}P_{\alpha}\,\sta{u_1,m_1}{1}\,\cdots\,\sta{u_N,m_N}{N}.
\end{equation}
Now, if $s$ is an integer, then $(-1)^{2sk_{\alpha}}=+1$ for any $k_{\alpha}$, and $\PSI{S}$ is totally symmetric (bosonic). If $s$ is a half-integer, then $(-1)^{2sk_{\alpha}}=-1$ for odd $k_{\alpha}$, and $+1$ for even $k_{\alpha}$, and $\PSI{S}$ is totally antisymmetric (fermionic).
\newpage
 
\noindent
{\textbf{6~~Generalization to non-product wave functions}

\noindent
Now we remove the restriction to wave functions of product form. The general normalized $N$-particle function then is
\begin{equation}
\Phi=\sum\nolimits_{\!\scriptsize{\begin{array}{l}r_1,\ldots,r_N\\s_1,\ldots,s_N\\t_1,\ldots,t_N \end{array}}}\!c_{r_1\cdots r_N s_1\cdots s_N t_1\cdots t_N}\;e^{\rmi m_{s1}\chi_{t1}}\psi^{(1)}\!(u_{r_1},m_{s_1})
\cdots e^{\rmi m_{sN}\chi_{tN}}\psi^{(N)}\!(u_{r_N},m_{s_N}),
\end{equation}
where the sum (or integral) over the $r$\sbl s and $t$\sbl s goes over a possibly infinite number of values, and the sum over the $s$'\sbl s goes over the $2s+1$ possible values of the spin component.

Permuting the parameter sets $\{u_{ri},m_{si}\}$ among the one-particle functions and permuting the angles $\{\chi_{ti}\}$ by means of rotations in every single term of the sum (17) now leads us to
\begin{displaymath}
\PHI{S}=\sum\nolimits_{\!\scriptsize{\begin{array}{l}r_1,\ldots,r_N\\s_1,\ldots,s_N\\t_1,\ldots,t_N \end{array}}}\!c_{r_1\cdots r_N s_1\cdots s_N t_1\cdots t_N}e^{\rmi (m_{s1}\chi_{t1}+\cdots+m_{rN}\chi_{tN})}
\end{displaymath}
\begin{equation}
\hspace{40pt}\times \frac{1}{\sqrt{N!}}\sum_{\alpha}(-1)^{2sk_{\alpha}}P_{\alpha}                           \psi^{(1)}\!(u_{r_1},m_{s_1})
\cdots \psi^{(N)}\!(u_{r_N},m_{s_N}),
\end{equation}
where we have used the results (15), (16) of the preceding section. Thus
the connection between spin and statistics is proved for general nonrelativistic $N$-particle functions.
\newline

\noindent
{\textbf{7~~The relativistic domain}

\noindent
The derivation of the spin-statistics connection presented so far evidently does not require relativity theory. Can it be extended into the relativistic domain? In Lorentz-invariant theory spin and orbital angular momentum are no longer separately conserved quantities, and the two are in general mixed up in a complicated way. There are however functions which are eigenfunctions of the spin-component operator only, with no admixture of orbital angular momentum: the helicity functions [44]. A helicity function describes a free particle with definite non-zero linear momentum and is an eigenfunction of the operator of the spin component with respect to an axis that is parallel or antiparallel to the direction of the particle's momentum. Thus we may replace the previously discussed eigenfunctions of the operator of the spin component along a fixed direction by the helicity functions. Helicities are invariant under ordinary rotations (involving spin and orbital part), and the rotation operators commute with the permutation operators, so we may express the momentum eigenfunctions which have their momenta in arbitrary directions by suitably rotated eigenfunctions with momenta in one common direction (cf. [44, pp. 407, 408]). For these functions we can define a common reference direction for the angles $\chi$, and then construct and add up the functions with the permuted parameters in the previously described way. This works not only for momentum eigenstates, i.e. plane waves, but also for linear superpositions of plane waves, i.e.~wave packets.
\newline

\noindent
{\textbf{Notes and References}}
\begin{enumerate}
\renewcommand{\labelenumi}{[\arabic{enumi}]}
\begin{sloppypar}

\item Fierz, M.: \"Uber die relativistische Theorie kr\"aftefreier Teilchen mit beliebigem Spin. Helv. Phys. Acta. \textbf{12}, 3-37 (1939). English translation of an excerpt in [4], pp. 285--300 

\item Pauli, W.: The connection between spin and statistics. Phys. Rev. \textbf{58}, 716--722 (1940) 

\item Jost, R.: Das Pauli-Prinzip und die Lorentz-Gruppe. In: Fierz, M.,  Weisskopf, V.F. (eds.) Theoretical Physics in the Twentieth Century, pp. 107--136. Interscience, New York (1960) 

\item  Duck, I., Sudarshan, E.C.G.: Pauli and the Spin-Statistics Theorem. World Scientific, Singapore (1997)

\item Feynman, R.P., Leighton, R.B., Sands, M.: The Feynman Lectures on Physics, vol. III, p. 4-3. Addison-Wesley, Reading (1965)

\item About half of these publications are accessible via the internet under arXiv.org/find [Title: spin AND statistics]; the others can be traced back from these

\item Hilborn, R.C.: Answer to Question \#7 [``The spin-statistics theorem,'' Dwight E. Neuenschwander, Am. J. Phys. 62 (6), 972 (1994)]. Am. J. Phys. \textbf{63}, 298--299 (1995) 

\item Duck, I., Sudarshan, E.C.G.: Toward an understanding of the spin-statistics theorem. Am. J. Phys. \textbf{66}, 284--303 (1998)

\item Romer, R.H.: The spin-statistics theorem. Am. J. Phys. \textbf{70}, 791 (2002) 

\item Morgan, J.A.: Spin and statistics in classical mechanics. arXiv:quant-ph/0401070 (Am. J. Phys. \textbf{72}, 1408--1417 (2004)) 

\item Broyles, A.A.: Derivation of the Pauli exchange principle. arXiv:quant-ph/9906046

\item Broyles, A.A.: Spin and statistics. Am. J. Phys. \textbf{44}, 340--343 (1976)  

\item Peshkin, M.: Reply to ``Non-relativistic proofs of the spin-statistics connection'', by Shaji and Sudarshan. arXiv:quant-ph/0402118 

\item Peshkin, M.: Reply to ``Comment on `Spin and statistics in nonrelativistic quantum mechanics: The spin-zero case' ''. Phys. Rev. A \textbf{68}, 046102 (2003) 

\item Peshkin, M.: Reply to ``No spin-statistics connection in nonrelativistic quantum mechanics''. arXiv:quant-ph/0306189 

\item Peshkin, M.: Spin and statistics in nonrelativistic quantum mechanics: The spin-zero case. Phys. Rev. A \textbf{67}, 042102 (2003) 

\item Peshkin, M.: On spin and statistics in quantum mechanics. arXiv:quant-ph/0207017

\item Morgan, J.A.: Demonstration of the spin-statistics connection in elementary quantum mechanics. arXiv:physics/0702058

\item  Kuckert, B.: Spin and statistics in nonrelativistic quantum mechanics. I. Phys. Lett. A \textbf{322}, 47--53 (2004) 

\item Donth, E.: Ein einfacher nichtrelativistischer Beweis des Spin-Statistik-Theorems und das Verh\"altnis von Geometrie und Physik in der Quantenmechanik. Wissenschaftliche Zeitschrift der Technischen Hochschule ,,Carl Schorlemmer`` Leuna-Merseburg \textbf{19}, 602--606 (1977) 

\item Donth, E.: Non-relativistic proof of the spin statistics theorem. Phys. Lett. A \textbf{32}, 209--210 (1970) 

\item Kuckert, B., Mund, J.: Spin \& statistics in nonrelativistic quantum mechanics, II. arXiv:quant-ph/0411197 (Ann. Physik (Leipzig) \textbf{14}, 309--311 (2005))  
 
\item Bacry, H.: Answer to Question \#7 [``The spin-statistics theorem,'' Dwight E. Neuenschwander, Am. J. Phys. 62 (6), 972 (1994)]. Am. J. Phys. \textbf{63}, 297--298 (1995) 

\item Bacry, H.: Introduction aux concepts de la physique statistique, pp. 198--200. Ellipses, Paris (1991) 

\item  Piron, C.: M\'ecanique quantique, Bases et applications, pp. 166--167. Presses polytechniques et universitaires romandes, Lausanne (1990)

\item Balachandran, A.P., Daughton, A., Gu, Z.-C., Sorkin, R.D., Marmo, G.,  Srivastava, A.M.: Spin-statistics theorems without relativity or field theory. Int. J. Modern Physics A \textbf{8}, 2993--3044 (1993) 

\item Feynman, R.P.: The reason for antiparticles. In: Feynman, R.P., Weinberg, S. (eds.) Elementary Particles and the Laws of Physics, pp. 1--59, especially pp. 56--59. Cambridge University Press, Cambridge (1987)
 
\item York, M.: Symmetrizing the symmetrization postulate. In: Hilborn, R.C., Tino, G.M. (eds.) Spin-Sta\-tis\-tics Connection and Commutation Relations, pp. 104--110. American Institute of Physics, Melville (2000). arXiv:quant-ph/0006101 

\item York, M.: Identity, geometry, permutation, and the spin-statistics theorem. arXiv:quant-ph/9908078

\item Jabs, A.: Quantum mechanics in terms of realism. arXiv:quant-ph/9606017 (Physics Essays \textbf{9}, 36--95, 354 (1996)). In this interpretation, which is opposed to the Copenhagen interpretation, the quantum objects are not point particles but extended objects, represented by the wave functions, like $\psi(\bfitr^{(i)}) \equiv \psi^{(i)}(\bfitr)$. The index $(i)$, rather than denoting the particle $i$ in the wave function, denotes the wave function itself. The vector $\bfitr^{(i)}$ does not mean the position of point particle~$i$ (where is it when its position is not being measured?) but the position variable of wave function $i$, and $|\psi(\bfitr^{(i)})|^2\mathrm{d}^3r$ is the probability that the wave function $i$ causes an effect about the point $\bfitr$. In either interpretation, realist or Copenhagen, these labels are needed to avoid self-interactions or to show the one-particle operators which wave function to operate on.  So, the construction of the proof presented here is independent of which of the interpretations is adopted, and the present paper sticks to the traditional formulation

\item Jabs, A.: An Interpretation of the Formalism of Quantum Mechanics in Terms of Epistemological Realism. arXiv:1212.4687 (Br. J. Philos. Sci. \textbf{43}, 405--421 (1992)) 
 
\item Cohen-Tannoudji, C., Diu, B., Lalo\"e, F.: Quantum Mechanics, vols.~I, II. Wiley, New York (1977)

\item Knopp, K.: Funktionentheorie, second part, pp. 90--91. de Gruyter, Berlin (1955). English translation by Bagemihl, F.: Theory of Functions, part II, pp. 101--103. Dover, New York (1996) 

\item Weyl, H.: The theory of groups and quantum mechanics, p. 184. Dover, New York (1950)

\item von Foerster, T.: Answer to Question \#7 [``The spin-statistics theorem,'' Dwight E. Neuenschwander, Am. J. Phys. 62 (6), 972 (1994)]. Am. J. Phys. \textbf{64} (5), 526 (1996) 

\item Gould, R.R.: Answer to Question \#7 [``The spin-statistics theorem,'' Dwight E. Neuenschwander, Am. J. Phys. 62 (6), 972 (1994)]. Am. J. Phys. \textbf{63} (2), 109 (1995) 

\item Penrose, R., Rindler, W.: Spinors and Space-Time, vol.~1, p. 43. Cambridge University Press, Cambridge (1984) 

\item Biedenharn, L.C., Louck, J.D.: Angular Momentum in Quantum Physics. Addison-Wesley, Reading (1981). Chapter~2 

\item Hartung, R.W.: Pauli principle in Euclidean geometry. Am. J. Phys. \textbf{47} (10), 900--910 (1979) 

\item Rieflin, E.: Some mechanisms related to Dirac's strings. Am. J. Phys. \textbf{47} (4), 378--381 (1979); and literature cited in [35--40]

\item Schr\"odinger, E.: Die Mehrdeutigkeit der Wellenfunktion. Ann. Physik (Leipzig) \textbf{32} (5), 49--55 (1938). Reprinted in: Schr\"odinger, E.:  Collected Papers, vol.~3, pp. 583--589. Verlag der \"Osterreichischen Akademie der Wissenschaften, Wien (1984)

\item Van Winter, C.: Orbital angular momentum and group representations. Ann. Phys. (New York) \textbf{47}, 232--274 (1968)

\item Altmann, S.L.: Rotations, Quaternions, and Double Groups, Dover, New York (2005). Chapter~10 

\item Jacob, M., Wick, G.C.: On the general theory of collisions for particles with spin. Ann. Phys. (New York) \textbf{7}, 404--428 (1959)

\end{sloppypar}
\end{enumerate}
\medskip
\hspace{5cm}
------------------------------
\end{document}